# Điều khiển ổn định robot di động phân tán qua mạng máy tính sử dụng bộ lọc dự đoán với quan sát quá khứ

## Stabilization control of networked mobile robot using past observation-based preditive filter

**Phùng Mạnh Dương, Nguyễn Thị Thanh Vân, Trần Thuận Hoàng, Trần Quang Vinh**
Trường ĐH Công nghệ, ĐHQGHN
e-Mail: duongpm@vnu.edu.vn

**Tóm tắt**

Bài báo trình bày vấn đề điều khiển ổn định cho hệ robot di động phân tán qua mạng máy tính chịu tác động của sự trì trễ. Một bộ ước lượng trạng thái mới có tên là bộ lọc dự đoán với quan sát quá khứ đã được xây dựng cho phép dự đoán trạng thái của hệ robot từ phép đo bị trì trễ. Dự đoán này kết hợp với luật điều khiển cho hệ truyền thống đảm bảo tính ổn định tiệm cận cho hệ robot phân tán. Các mô phỏng với tham số lấy từ hệ thực đã được tiến hành và kết quả khẳng định tính đúng đắn cũng như khả năng áp dụng của giải thuật vào các hệ thực.

**Abstract:** This paper addresses the stabilization control problem for networked mobile robot subject to communication delay. A new state estimation filter namely past observation-based predictive filter is developed. This filter enables the prediction of system state from delayed measurement. The state estimator combined with developed control laws ensures the asymptotic stability of the networked system. Simulations with parameters extracted from a real robot system were conducted and results confirmed the correctness as well as applicability of proposed approach.

**Chữ viết tắt**
PO-PF      Past observation-based Predective filter

## 1. Mở đầu

Điều khiển robot phân tán qua mạng máy tính đang nhận được nhiều sự quan tâm nghiên cứu gần đây nhờ khả năng mở ra những ứng dụng mới cho cuộc sống hiện đại như hệ thống mổ từ xa, phòng thí nghiệm ảo hay cứu hộ trong môi trường phóng xạ. Nếu như những nghiên cứu ban đầu cố gắng trả lời câu hỏi làm thế nào để điều khiển được một robot qua mạng Internet [1, 2] thì các nghiên cứu gần đây lại tập trung giải quyết các thách thức cụ thể như khắc phục độ trì trễ, tối ưu băng thông, lựa chọn giao thức, định vị trong điều kiện nhiễu… [3-6]. Trong số đó, vấn đề điều khiển ổn định cần nhận được những khảo sát và nghiên cứu chi tiết.

Điều khiển ổn định là vấn đề cơ bản trong lý thuyết điều khiển nói chung và trong robot di động nói riêng. Việc di chuyển một cách ổn định từ một điểm khởi phát đến điểm đích là cơ sở cho sự vận hành hiệu quả của toàn hệ thống và là nền tảng để xây dựng các ứng dụng thực tiễn. Từ ý nghĩa đó, nhiều nghiên cứu đã được thực hiện và bài toán điều khiển ổn định robot di động tập trung truyền thống thực tế đã được giải quyết cả về mặt thực nghiệm lẫn lý thuyết [7-9]. Tuy nhiên, hệ thống robot di động khi phân tán qua mạng máy tính có những điểm khác biệt. Đó là sự tác động của các tham số mạng như độ trì trễ, độ mất mát dữ liệu, sự sai lệch thứ tự dữ liệu truyền hay sự giới hạn băng thông cho phép… lên tín hiệu điều khiển và phản hồi. Xét từ góc độ điều khiển học, những tác động này gây ra sự không chính xác trong ước lượng trạng thái và có thể làm giảm đáng kể hiệu năng của hệ thống. Do đó, một số đề xuất cho vấn đề này đã được đưa ra với những ưu và nhược điểm khác nhau. Trong [10], A. Ray đề xuất việc sử dụng một bộ đệm thời gian có giá trị dài hơn độ trì trễ trong trường hợp xấu nhất để đưa hệ thống trở về bất biến với thời gian và từ đó sử dụng lý thuyết điều khiển cổ điển. Trong [11], các tác giả đã mô hình hóa thời gian trễ từ cảm biến tới bộ điều khiển và từ bộ điều khiển tới cơ cấu chấp hành sử dụng chuỗi Markov. Trong [12], Wenshan Hu đề xuất một mô hình điều khiển dự đoán dựa trên dữ liệu phép đo về thời gian truyền (Round-trip time). Trong tiếp cận này, một tập hợp các tín hiệu điều khiển cho tất cả các khả năng của thời gian trễ được đóng gói đồng thời trong một gói tin và gửi tới hệ chấp hành. Ở phía hệ chấp hành, tín hiệu điều khiển phù hợp sẽ được lựa chọn dựa trên giá trị trì trễ đo được. Tuy nhiên, các phương pháp trên đều giả sử rằng hệ thống là tuyến tính và việc mở rộng chúng cho các hệ phi tuyến như robot di động cần nhiều thời gian nghiên cứu.

Trong một hướng tiếp cận khác, Nielsen đã giải quyết tương đối hoàn chỉnh bài toán điều khiển khiển ổn định cho các hệ phi tuyến với điều kiện thời gian trễ nhỏ hơn chu kỳ lấy mẫu [13]. Trong trường hợp trễ lớn hơn, Wargui đã đề xuất sử dụng một bộ ước lượng để dự đoán trạng thái của hệ thống tại thời điểm tương lai mà tín hiệu điều khiển hiện tại sẽ tới hệ chấp hành [14]. Từ đó, thay vì tạo tín hiệu cho phản hồi hiện tại, bộ điều khiển gửi tín hiệu điều khiển cho trạng thái dự đoán. Hướng tiếp cận này khả thi và hiệu quả tuy nhiên phụ thuộc nhiều vào sự chính xác trong ước lượng trạng thái.

Bài báo này trình bày vấn đề ổn định cho robot di động khi điều khiển phân tán qua mạng máy tính trong đó hệ thống được giả thiết là chỉ chịu tác động của sự trì trễ. Các ảnh hưởng khác của mạng truyền thông như độ mất mát dữ liệu, sự sai khác thứ tự gói tin, giao thức truyền tải … sẽ được khảo sát trong các nghiên cứu tiếp theo. Từ giới hạn này, hướng tiếp cận của chúng tôi là sử dụng một bộ ước lượng trạng thái tương tự [14]. Tuy nhiên, cấu trúc của bộ ước lượng là hoàn toàn khác trong đó chúng tôi đề xuất một bộ lọc mới có tên là *bộ lọc dự đoán với quan sát quá khứ PO-PF* (past observation-based predictive filter). Bộ lọc này kết hợp dữ liệu của mô hình động học hệ thống, tín hiệu điều khiển lối vào và phép đo phản hồi đã bị trì trễ để dự đoán một cách tối ưu (về mặt thống kê) trạng thái của hệ robot. Các mô phỏng thực nghiệm đã được thực hiện và kết quả đã khẳng định tính đúng đắn của đề xuất.

Bài báo được trình bày theo cấu trúc như sau. Phần II giới thiệu mô hình hệ thống và đặt vấn đề bài toán. Phần III trình bày bộ ước lượng với bộ lọc PO-PF. Phần IV trình bày cài đặt chi tiết hệ mô phỏng và kết quả. Bài báo kết thúc với những thảo luận và đánh giá về phương pháp đã đề xuất.

## 2. Đặt vấn đề và mô hình hệ thống

Phần này trình bày tóm tắt vấn đề điều khiển ổn định cho robot di động không phân tán. Trên cơ sở đó, hệ robot phân tán được mô hình hóa và phân tích để đưa đến hướng tiếp cận giải quyết bài toán ổn định.

### 2.1 Điều khiển ổn định robot di động không phân tán

Báo cáo tập trung vào loại robot di động có hai bánh vi sai với ràng buộc không khả tích (non-holonomic). Mô hình robot được trình bày trong hình 1, trong đó, $(X_G, Y_G)$ biểu diễn hệ tọa độ toàn cục, $(X_R, Y_R)$ biểu diễn hệ tọa độ cục bộ gắn liền với robot, $R$ ký hiệu bán kính bánh xe và $L$ là khoảng cách giữa hai bánh.

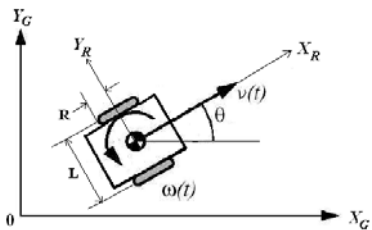

**H. 1**  *Mô hình robot di động hai bánh vi sai*

Mô hình động học của robot được mô tả như sau:
$$\dot{x} = v\cos\theta$$
$$\dot{y} = v\sin\theta \quad (1)$$
$$\dot{\theta} = \omega$$

Trong đó $(x, y)$ là tọa độ của robot, $\theta$ là hướng của robot, $v$ và $\omega$ lần lượt là vận tốc dài và vận tốc góc của robot: $v = (\omega_L + \omega_R)/2$, $\omega = (\omega_L - \omega_R)/L$ với $\omega_L$ và $\omega_R$ lần lượt là vận tốc góc của bánh trái và bánh phải. Sử dụng lý thuyết Lyapunov, Aicardi [7] đã chứng minh rằng hệ (1) ổn định tiệm cận với luật điều khiển sau:

$$v = (\gamma \cos\alpha)\rho$$
$$w = \lambda\alpha + \gamma\frac{\cos\alpha\sin\alpha}{\alpha}(\alpha + h\phi) \quad (2)$$

Trong đó $\gamma$, $\lambda$ và $h$ là các tham số dương; gọi $O_1X_1Y_1$ và $O_2X_2Y_2$ là hệ tọa độ gắn với robot tại điểm đầu và điểm đích, khi đó $\rho$ được định nghĩa là khoảng cách giữa $O_1$ và $O_2$, $\phi$ là góc tạo bởi vecto nối $O_1$ và $O_2$ và vecto nối $O_2$ và $x_2$, $\alpha$ là góc tạo bởi vecto nối $O_1$ và $O_2$ và vecto nối $O_1$ và $x_1$ (Hình 2).

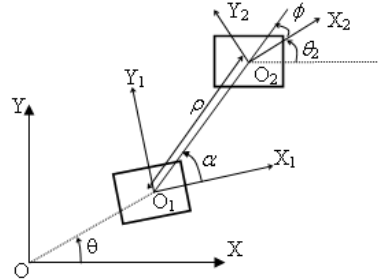

**H. 2**  *Robot trong không gian biến dẫn đường*

Khai triển Taylor, mô hình rời rạc của (1) có dạng:
$$x_{k+1} = x_k + T_s v_k \cos\theta_k$$
$$y_{k+1} = y_k + T_s v_k \sin\theta_k \quad (3)$$
$$\theta_{k+1} = \theta_k + T_s \omega_k$$

và luật điều khiển ổn định trong miền rời rạc có dạng:
$$v_k = (\gamma \cos\alpha_k)\rho_k$$
$$w_k = \lambda\alpha_k + \gamma\frac{\cos\alpha_k \sin\alpha_k}{\alpha_k}(\alpha_k + h\phi_k) \quad (4)$$

Mô hình rời rạc (3) và luật điều khiển (4) là cơ sở cho thuật toán điều khiển ổn định được trình bày tiếp theo.

### 2.2 Mô hình điều khiển robot qua mạng máy tính

Xét hệ robot điều khiển qua mạng máy tính. Hệ thống trở thành phân tán và hoạt động của hệ bị tác động bởi các tham số mạng như thời gian trì trễ, sự mất mát dữ liệu hay băng thông cho phép. Trong số các tham số này, độ trì trễ có ảnh hưởng chính và sẽ được đề cập trong bài báo này. Các tham số khác sẽ được khảo sát trong các nghiên cứu tiếp theo. Hình 3 biểu diễn hệ robot khi được điều khiển qua mạng máy tính với $n$ và $m$ lần lượt là độ trì trễ của tín hiệu điều khiển và tín hiệu đo.

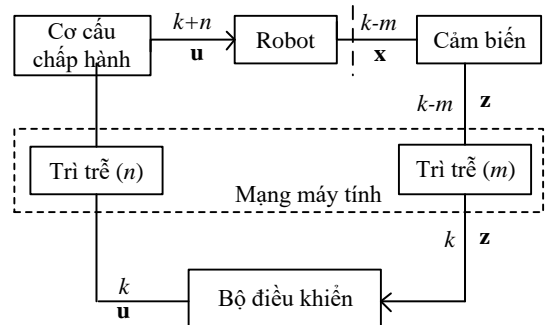

**H. 3**  *Mô hình hệ thống điều khiển robot qua mạng máy tính*

Có thể thấy rằng, do sự trì trễ, tại thời điểm $k$, bộ điều khiển chỉ nhận được tín hiệu đo tại thời điểm $k-m$ thay vì $k$. Tương tự, tín hiệu điều khiển gửi đi tại $k$ sẽ chỉ tới cơ cấu chấp hành vào thời điểm *k+n*. Hệ thống trở thành có nhớ và luật điều khiển (4) không còn đảm bảo hệ thống (3) ổn định tiệm cận. Tuy nhiên, nếu ta xây dựng được một bộ ước lượng trạng thái sao cho tại thời điểm *k*, với dữ liệu phép đo tại thời điểm *k-m* và mô hình hệ thống, có thể dự đoán được trạng thái của hệ thống tại thời điểm *k+n* để từ đó gửi tín hiệu điều khiển **u**$_{k+n}$ thì hệ thống sẽ lại ổn định tiệm cận [14].

Một cách lượng hóa, gọi trạng thái của robot bao gồm vị trí và hướng được biểu diễn bởi vecto $\mathbf{x} = [x\ y\ \theta]^T$. Trạng thái này có thể được đo bởi phép đo **z**. Phép đo này được biểu diễn bằng một hàm phi tuyến, *h*, của trạng thái robot và nhiễu phép đo, **v**. Kí hiệu hàm (3) là *f*, với vecto lối vào **u** và nhiễu lối vào **w**, mô hình robot khi điều khiển qua mạng được biểu diễn trong không gian trạng thái như sau:

$$\begin{aligned}\mathbf{x}_k &= f_{k-1}(\mathbf{x}_{k-1}, \mathbf{u}_{k-n-1}, \mathbf{w}_{k-1}) \\ \mathbf{z}_k &= h_{k-m}(\mathbf{x}_{k-m}, \mathbf{v}_{k-m})\end{aligned} \quad (5)$$

Từ (5), chúng tôi đề xuất thuật toán cho phép, tại thời điểm *k*, có thể ước lượng trạng thái tại *k+n*, $\hat{\mathbf{x}}(k+n | k-m)$, từ dữ liệu đo và tín hiệu điều khiển bị trễ. Trạng thái ước lượng này kết hợp với (4) sẽ đảm bảo điều khiển ổn định hệ thống robot di động qua mạng máy tính.

## 3. Thuật toán ước lượng trạng thái và điều khiển ổn định

Phần này trình bày giải thuật dự đoán trạng thái của hệ thống trong điều kiện cả tín hiệu điều khiển và phép đo phản hồi đều bị trì trễ. Thuật toán có tên *bộ lọc dự đoán từ quan sát quá khứ PO-PF* được xây dựng trên cơ sở lý thuyết về bộ lọc Kalman. Trong phần này, bộ lọc Kalman sẽ được trình bày ngắn gọn làm cơ sở để xây dựng bộ lọc PO-PF cho hệ tuyến tính. Bộ lọc PO-PF sau đó được mở rộng để áp dụng cho các hệ phi tuyến bao gồm robot di động.

### 3.1 Bộ lọc Kalman

Bộ lọc Kalman theo định nghĩa là một tập hợp các phương trình toán học đệ quy để ước lượng trạng thái của một quá trình, sao cho tối thiểu trung bình của bình phương sai số. Xét một hệ thống tuyến tính rời rạc biểu diễn trong không gian trạng thái trong đó cả tín hiệu đầu vào và phép đo đều bị tác động bởi nhiễu trắng phân bố Gauss như sau:

$$\begin{aligned}\mathbf{x}_k &= A_{k-1}\mathbf{x}_{k-1} + B_{k-1}\mathbf{u}_{k-1} + \mathbf{w}_{k-1} \\ \mathbf{z}_k &= H_k\mathbf{x}_k + \mathbf{v}_k\end{aligned} \quad (6)$$

Khi đó, bộ lọc Kalman được tính toán qua hai bước dự đoán và hiệu chỉnh như sau:

- Pha dự đoán:

$$\begin{aligned}\hat{\mathbf{x}}_k^- &= A_{k-1}\hat{\mathbf{x}}_{k-1}^+ + B_{k-1}\mathbf{u}_{k-1} \\ P_k^- &= A_{k-1}P_{k-1}^+ A_{k-1}^T + Q_{k-1}\end{aligned} \quad (7)$$

Trong đó $\hat{\mathbf{x}}_k^- \in \Re^n$ là tiền ước lượng tại bước *k*, $P_k^-$ là ma trận hiệp phương sai của sai số ước lượng và $Q_{k-1}$ là ma trận hiệp phương sai của nhiễu lối vào.

- Pha hiệu chỉnh:

$$\begin{aligned}K_k &= P_k^- H_k^T [H_k P_k^- H_k^T + R_k]^{-1} \\ \hat{\mathbf{x}}_k^+ &= \hat{\mathbf{x}}_k^- + K_k[\mathbf{z}_k - H_k\hat{\mathbf{x}}_k^-] \\ P_k^+ &= [I - K_k H_k]P_k^-\end{aligned} \quad (8)$$

Trong đó $\hat{\mathbf{x}}_k^+ \in \Re^n$ là hậu ước lượng tại bước *k* biết dữ liệu đo $\mathbf{z}_k$, $K_k$ là hệ số khuếch đại Kalman và $R_k$ là ma trận hiệp phương sai của nhiễu đo.

### 3.2 Bộ lọc dự đoán từ quan sát quá khứ PO-PF

Bộ lọc dự đoán trước hết ước lượng trạng thái hiện tại của hệ thống từ mô hình và phép đo bị trễ (5). Ước lượng này sau đó được ngoại suy tới thời điểm cần đặt tín hiệu.

Từ hình 2 và phương trình (5), trạng thái tại thời điểm *k* phản ánh tác động của lối vào tại *k-n-1*. Pha dự đoán của bộ lọc Kalman có thể viết lại như sau:

$$\hat{\mathbf{x}}_k^- = A_{k-1}\hat{\mathbf{x}}_{k-1}^+ + B_{k-n-1}\mathbf{u}_{k-n-1} \quad (9)$$

Gọi $\mathbf{z}_k^* = H_s\mathbf{x}_s + \mathbf{v}_s$ là phép đo bị trễ *m* chu kỳ tới bộ ước lượng trạng thái vào thời điểm *k*. Phép đo này thực chất phản ánh trạng thái của hệ thống tại thời điểm quá khứ *s* hơn là tại *k*. Việc kết hợp phép đo $\mathbf{z}_k^*$ trực tiếp vào phương trình hiệu chỉnh (8) tại thời điểm *k* do đó không thể thực hiện được. Tuy nhiên, nếu ta xem sự thay đổi giá trị của phép đo từ thời điểm *s* tới *k* như là sự sai khác giữa các tiền ước lượng của pha dự đoán, phép đo hiện tại $\mathbf{z}_k^{pre}$ khi đó có thể ngoại suy từ phép đo bị trễ $\mathbf{z}_k^*$ như sau:

$$\begin{aligned}\mathbf{z}_k^{pre} &= \mathbf{z}_k^* + (H_k\hat{\mathbf{x}}_k^- - H_s\hat{\mathbf{x}}_s^-) \\ &= H_s\mathbf{x}_s + \mathbf{v}_s + H_k\hat{\mathbf{x}}_k^- - H_s\hat{\mathbf{x}}_s^- \\ &= H_k\mathbf{x}_k + H_k\tilde{\mathbf{x}}_k^- - H_s\tilde{\mathbf{x}}_s^- + \mathbf{v}_s \\ &= H_k\mathbf{x}_k + \mathbf{v}_k^{pre}\end{aligned} \quad (10)$$

Trong đó, sai số ước lượng $\tilde{\mathbf{x}}_k^- = \hat{\mathbf{x}}_k^- - \mathbf{x}_k$. Kết hợp phép đo dự đoán $\mathbf{z}_k^{pre}$ vào phương trình hiệu chỉnh (8) cho:

$$\begin{aligned}\hat{\mathbf{x}}_k^+ &= \hat{\mathbf{x}}_k^- + K_k[\mathbf{z}_k^{pre} - H_k\hat{\mathbf{x}}_k^-] \\ &= \hat{\mathbf{x}}_k^- + K_k[\mathbf{z}_k^* + H_k\hat{\mathbf{x}}_k^- - H_s\hat{\mathbf{x}}_s^- - H_k\hat{\mathbf{x}}_k^-] \\ &= \hat{\mathbf{x}}_k^- + K_k[\mathbf{z}_k^* - H_s\hat{\mathbf{x}}_s^-]\end{aligned} \quad (11)$$

Để đảm bảo sự tối ưu trong phép kết hợp phép đo dự đoán, ta cần tính lại các hệ số kalman $K_k$ và ma trận hiệp phương sai sai số ước lượng $P_k^+$. Giả sử phương trình (11) được thực thi với một giá trị tùy ý của $K_k$, sai số ước lượng, $\tilde{\mathbf{x}}_k^+$, trở thành :

$$\tilde{\mathbf{x}}_k^+ = \hat{\mathbf{x}}_k^+ - \mathbf{x}_k = (I - K_k H_k)\tilde{\mathbf{x}}_k^- + K_k\mathbf{v}_k^{pre} \quad (12)$$

Ma trận hiệp phương sai được tính bởi:

$$\begin{aligned}P_k^+ &= E\{\tilde{\mathbf{x}}_k^+ \tilde{\mathbf{x}}_k^{+T}\} \\ &= (I - K_k H_k)P_k^-(I - K_k H_k)^T + (I - K_k H_k)E\{\tilde{\mathbf{x}}_k^- \mathbf{v}_k^{preT}\}K_k^T \\ &\quad + K_k E\{\mathbf{v}_k^{pre}\tilde{\mathbf{x}}_k^{-T}\}(I - K_k H_k)^T + K_k E\{\mathbf{v}_k^{pre}\mathbf{v}_k^{preT}\}K_k^T\end{aligned}$$

$$(13)$$

Từ (10) và tính chất độc lập giữa $\tilde{\mathbf{x}}_k^-$ và $\mathbf{v}_k$, các hiệp phương sai trong (13) có thể tính được như sau:

$$E\{\tilde{\mathbf{x}}_k^- \mathbf{v}_k^{pre^T}\} = P_k^- H_k^T - M^T H_s^T \quad (14)$$

$$E\{\mathbf{v}_k^{pre} \mathbf{v}_k^{pre^T}\} = R_s + H_k P_k^- H_k^T + H_s P_s^- H_s^T \\ - H_s M H_k^T - H_k M^T H_s^T \quad (15)$$

Trong đó $M = E\{\tilde{\mathbf{x}}_s^- \tilde{\mathbf{x}}_k^{-T}\}$. Thay (14) và (15) vào (13) ta thu được:

$$P_k^+ = P_k^- - M^T H_s^T K_k^T - K_k H_s M \\ + K_k H_s P_s^- H_s^T K_k^T + K_k R_s K_k^T \quad (16)$$

Trong lý thuyết về bộ lọc Kalman, ma trận $K_k$ được chọn sao cho tối thiểu hiệp phương sai của sai số hậu ước lượng [15]. Phép tối thiểu này được thực hiện bằng cách lấy đạo hàm của vết (trace) của hiệp phương sai sai số ước lượng với $K_k$, đặt đạo hàm này bằng 0 và từ đó thu được $K_k$. Áp dụng quy trình trên vào hệ thống của ta, thu được:

$$\frac{\partial tr(P_k^+)}{\partial K_k} = -2H_s M + 2H_s P_s^- H_s^T K_k^T + (R_s + R_s^T) K_k^T = 0$$

$$\Leftrightarrow K_k = M^T H_s^T [H_s P_s^- H_s^T + R_s]^{-1} \quad (17)$$

Thay (17) vào (16) thu được phương trình của $P_k^+$:

$$P_k^+ = P_k^- - K_k H_s M \quad (18)$$

Để tính $M$, ta cần tính tiền ước lượng tại thời điểm $k$ từ ước lượng tại thời điểm $s$. Từ phương trình dự đoán (9) và phương trình hiệu chỉnh (11), $\tilde{\mathbf{x}}^-$ có dạng:

$$\tilde{\mathbf{x}}_k^- = \hat{\mathbf{x}}_k^- - \mathbf{x}_k = A_{k-1}\tilde{\mathbf{x}}_{k-1}^+ - \mathbf{w}_{k-1} \\ = A_{k-1}[(I - K_{k-1}H_{k-1})\tilde{\mathbf{x}}_{k-1}^- + K_{k-1}\mathbf{v}_{k-1}] - \mathbf{w}_{k-1} \quad (19)$$

Sau $m$ chu kỳ tính từ thời điểm $s$ tới $k$, $\tilde{\mathbf{x}}^-$ trở thành:

$$\tilde{\mathbf{x}}_k^- = M_* \tilde{\mathbf{x}}_s^- + f_1(\mathbf{w}_s...\mathbf{w}_{k-1}) + f_2(\mathbf{v}_{s+1}...\mathbf{v}_k) \quad (20)$$

Trong đó:
$$M_* = \prod_{i=1}^{m} A_{k-i}(I - K_{k-i}H_{k-i}) \quad (21)$$

$f_1$ và $f_2$ là các hàm của nhiễu $\mathbf{w}$ và $\mathbf{v}$. Từ (20) và sự độc lập giữa $\tilde{\mathbf{x}}$ với các nhiễu $\mathbf{v}$, $\mathbf{w}$, ta thu được:

$$M = E\{\tilde{\mathbf{x}}_s^- \tilde{\mathbf{x}}_k^{-T}\} = P_s^- M_*^T \quad (22)$$

Thay (22) vào (18) và (17) thu được:

$$P_k^+ = P_k^- - K_k H_k P_s^- M_*^T \quad (23)$$

và $\quad K_k = M_* P_s^- H_s^T [H_s P_s^- H_s^T + R_s]^{-1} = M_* K_s^* \quad (24)$

Trong đó $K_s^*$ là hệ số Kalman tại thời điểm $s$ của bộ lọc Kalman chuẩn (8).

Từ (24), có thể nhận thấy rằng việc cập nhật phép đo bị trễ vào tính toán Kalman hiện tại $k$ thực chất được thực hiện bình thường như tại thời điểm $s$ nhưng hệ số Kalman cần thay đổi một lượng nhân $M_*$. Hệ số này phản ánh sự tương quan của phép đo quá khứ tại thời điểm $s$ với trạng thái hiện tại $k$.

**3.3 Mở rộng bộ lọc PO-PF cho hệ robot phi tuyến phân tán qua mạng**

Bộ lọc PO-PF đã xây dựng ở trên có thể ứng dụng cho hệ điều khiển phân tán nhưng đòi hỏi hệ phải tuyến tính. Phần này trình bày việc mở rộng bộ lọc PO-PF cho hệ phi tuyến. Ý tưởng cho việc mở rộng dựa trên phương pháp xây dựng bộ lọc Kalman mở rộng. Đó là sự tuyến tính hóa hệ phi tuyến quanh các điểm ước lượng trước.

Thực hiện khai triển Taylor cho phương trình trạng thái tại điểm $(\hat{\mathbf{x}}_{k-1}^+, \mathbf{u}_{k-1}, 0)$ thu được:

$$\mathbf{x}_k = f_{k-1}(\hat{\mathbf{x}}_{k-1}^+, \mathbf{u}_{k-1}, 0) + \frac{\partial f_{k-1}}{\partial \mathbf{x}}\bigg|_{(\hat{\mathbf{x}}_{k-1}^+, \mathbf{u}_{k-1}, 0)} (\mathbf{x}_{k-1} - \hat{\mathbf{x}}_{k-1}^+) \\ + \frac{\partial f_{k-1}}{\partial \mathbf{w}}\bigg|_{(\hat{\mathbf{x}}_{k-1}^+, \mathbf{u}_{k-1}, 0)} \mathbf{w}_{k-1} \quad (25) \\ = f_{k-1}(\hat{\mathbf{x}}_{k-1}^+, \mathbf{u}_{k-1}, 0) + A_{k-1}(\mathbf{x}_{k-1} - \hat{\mathbf{x}}_{k-1}^+) + W_{k-1}\mathbf{w}_{k-1} \\ = A_{k-1}\mathbf{x}_{k-1} + [f_{k-1}(\hat{\mathbf{x}}_{k-1}^+, \mathbf{u}_{k-1}, 0) - A_{k-1}\hat{\mathbf{x}}_{k-1}^+] + W_{k-1}\mathbf{w}_{k-1} \\ = A_{k-1}\mathbf{x}_{k-1} + \tilde{\mathbf{u}}_{k-1} + \tilde{\mathbf{w}}_{k-1}$$

Trong đó $A_{k-1}$, $W_{k-1}$, $\tilde{\mathbf{u}}_{k-1}$, $\tilde{\mathbf{w}}_{k-1}$ xác định bởi phương trình trên. Tương tự, tuyến tính hóa hàm đo tại $\mathbf{x}_k = \hat{\mathbf{x}}_k^-$ và $\mathbf{v}_k = 0$ thu được:

$$\mathbf{z}_k = h_k(\hat{\mathbf{x}}_k^-, 0) + \frac{\partial h_k}{\partial \mathbf{x}}\bigg|_{(\hat{\mathbf{x}}_k^-, 0)} (\mathbf{x}_k - \hat{\mathbf{x}}_k^-) + \frac{\partial h_k}{\partial \mathbf{v}}\bigg|_{(\hat{\mathbf{x}}_k^-, 0)} \mathbf{v}_k \\ = h_k(\hat{\mathbf{x}}_k^-, 0) + H_k(\mathbf{x}_k - \hat{\mathbf{x}}_k^-) + V_k \mathbf{v}_k \quad (26) \\ = H_k \mathbf{x}_k + [h_k(\hat{\mathbf{x}}_k^-, 0) - H_k \hat{\mathbf{x}}_k^-] + V_k \mathbf{v}_k \\ = H_k \mathbf{x}_k + \tilde{\mathbf{z}}_k + \tilde{\mathbf{v}}_k$$

Trong đó $H_k$, $V_k$, $\tilde{\mathbf{z}}_k$, $\tilde{\mathbf{v}}_k$ xác định bởi phương trình trên. Phương trình hệ thống (25) và phép đo (26) bây giờ trở thành tuyến tính. Áp dụng bộ lọc PO-PF vào các phương trình này kết hợp với pha ngoại suy thu được bộ lọc PO-PF hoàn chỉnh cho hệ robot phi tuyến phân tán quang mạng máy tính như sau:

- Pha dự đoán:

$$\hat{\mathbf{x}}_k^- = f_{k-1}(\hat{\mathbf{x}}_{k-1}^+, \mathbf{u}_{k-n-1}, \mathbf{0}) \\ P_k^- = A_k P_{k-1}^+ A_k^T + W_k Q_{k-1} W_k^T \quad (27)$$

- Pha hiệu chỉnh:

$$M_* = \prod_{i=1}^{m} A_{k-i}(I - K_{k-i}H_{k-i}) \\ K_k = M_* P_s^- H_s^T (H_s P_s^- H_s^T + V_s R_s V_s^T)^{-1} \quad (28) \\ \hat{\mathbf{x}}_k^+ = \hat{\mathbf{x}}_k^- + K_k [\mathbf{z}_k^* - h(\hat{\mathbf{x}}_k^-, \mathbf{0})] \\ P_k^+ = P_k^- - K_k H_k P_s^- M_*^T$$

- Pha ngoại suy:

$$\hat{\mathbf{x}}_{k+n}^- = f_{k+n-1}(\hat{\mathbf{x}}_{k+n-1}^-, \mathbf{u}_{k+n-1}, \mathbf{0}) \quad (29)$$

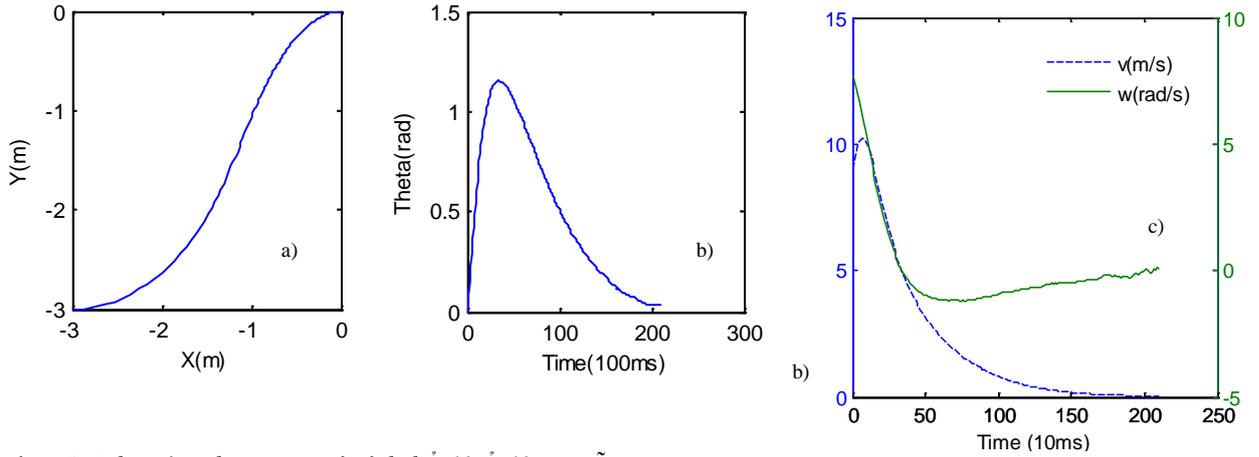

**H. 4**   *Quỹ đạo của robot trong quá trình điều khiển không trễ.*
   *a) Quỹ đạo trong mặt phẳng chuyển động OXY   b) Biến thiên hướng của robot   c) Biến thiên vận tốc dài và vận tốc góc*

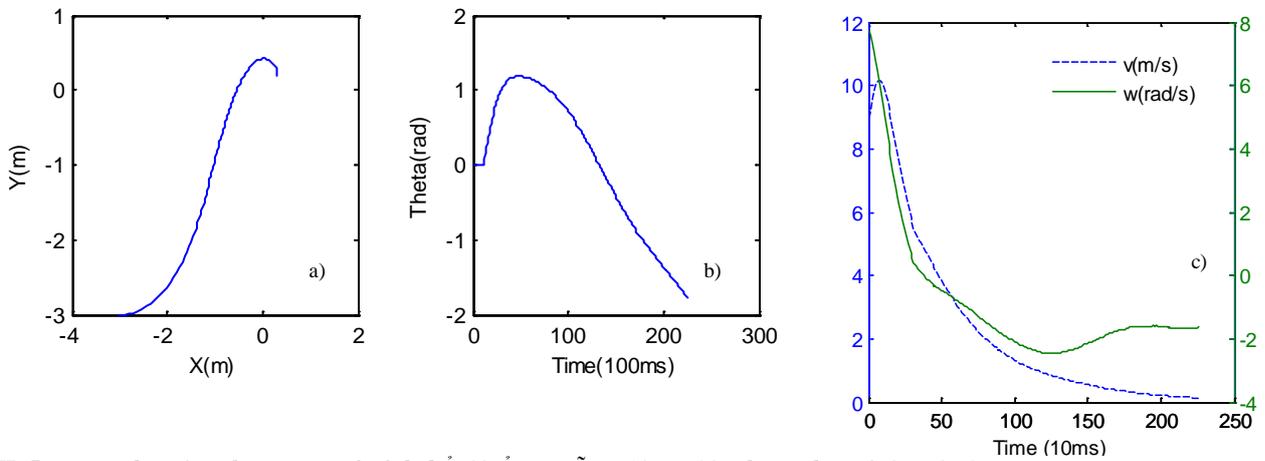

**H. 5**   *Quỹ đạo của robot trong quá trình điều khiển có trễ (n=10, m=20) nhưng chưa sử dụng bộ lọc PO-PF.*
   *a) Quỹ đạo trong mặt phẳng chuyển động OXY   b) Biến thiên hướng của robot   c) Biến thiên vận tốc dài và vận tốc góc*

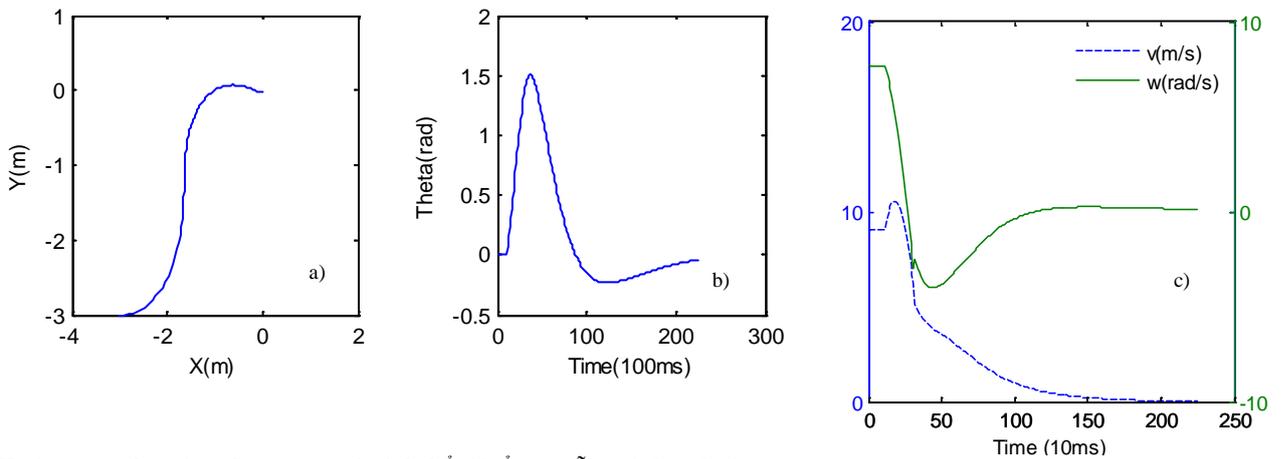

**H. 6**   *Quỹ đạo của robot trong quá trình điều khiển có trễ có sử dụng bộ lọc PO-PF.*
   *a) Quỹ đạo trong mặt phẳng chuyển động OXY   b) Biến thiên hướng của robot   c) Biến thiên vận tốc dài và vận tốc góc*

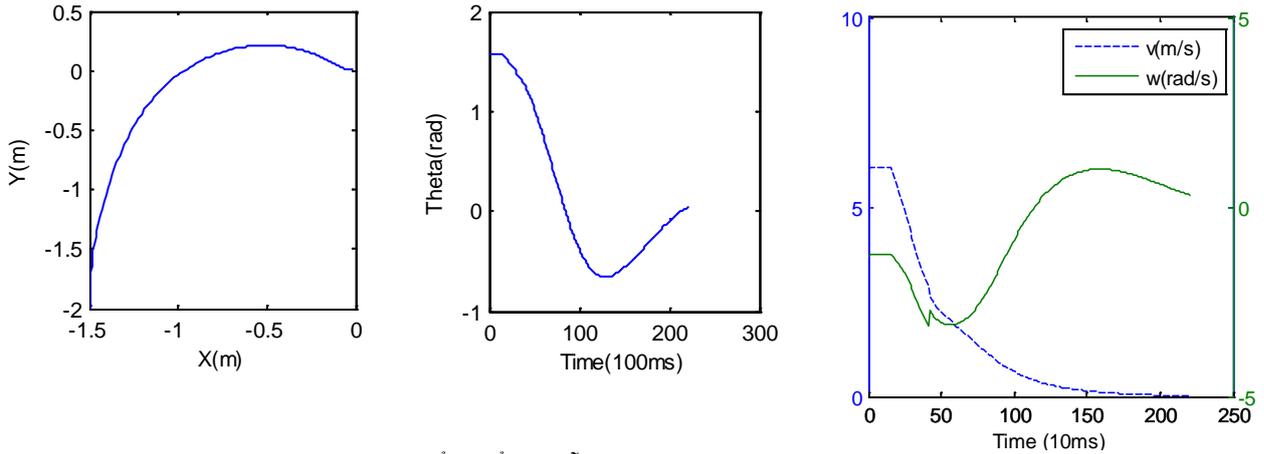

**H. 7** *Quỹ đạo của robot trong quá trình điều khiển có trễ (n=15, m=25) có sử dụng bộ lọc PO-PF.*
*a) Quỹ đạo trong mặt phẳng chuyển động OXY   b) Biến thiên hướng của robot   c) Biến thiên vận tốc dài và vận tốc góc*

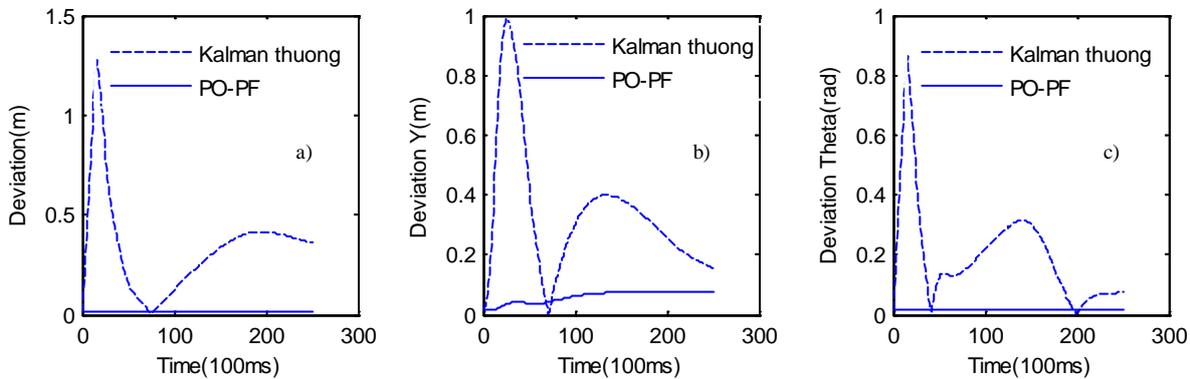

**H. 8** *So sánh độ lệch của đường ước lượng với đường thực cho bộ lọc Kalman và bộ lọc PO-PF.*
*a) Độ lệch theo phương X   b) Độ lệch theo phương Y   c) Độ lệch hướng θ*

## 4. Mô phỏng

Để đánh giá hiệu quả và khả năng ứng dụng của bộ lọc PO-PF kết hợp với luật điều khiển (4) cho bài toán điều khiển ổn định robot di động phân tán, chúng tôi đã tiến hành mô phỏng giải thuật trên Matlab.

### 4.1 Cài đặt mô phỏng

Trong thiết lập chương trình mô phỏng, robot là loại hai bánh vi sai với mô hình động học mô tả trong phần II. Bán kính của bánh xe là 0.05m và khoảng cách giữa các bánh là 0.6m. Thời gian lấy mẫu là 10ms. Các tham số cho bộ điều khiển (4) được lựa chọn như sau: $\lambda=6$, $h=1$, $\gamma=3$. Ma trận hiệp phương sai của nhiễu lối vào $Q$ và nhiễu phép đo $R$ được lựa chọn như sau:

$$Q_k = \begin{bmatrix} 0.01 & 0 \\ 0 & 0.01 \end{bmatrix} \quad R_k = \begin{bmatrix} 0.01 & 0 & 0 \\ 0 & 0.01 & 0 \\ 0 & 0 & 0.018 \end{bmatrix} \quad (30)$$

Các tham số này được thiết lập trên cơ sở một hệ thống robot thực đã được xây dựng tại phòng nghiên cứu của chúng tôi [16]. Các tham số còn lại cần thiết cho việc cài đặt mô phỏng được tính từ mô hình trạng thái robot (5):

$$A_{k+1} = \left.\frac{\partial f_k}{\partial \mathbf{x}}\right|_{(\hat{\mathbf{x}}_k^+, \mathbf{u}_k, \mathbf{0})} = \begin{bmatrix} 1 & 0 & -T_s v_c \sin\hat{\theta}_k^+ \\ 0 & 1 & T_s v_c \cos\hat{\theta}_k^+ \\ 0 & 0 & 1 \end{bmatrix} \quad (31)$$

$$W_{k+1} = \left.\frac{\partial f_k}{\partial \mathbf{w}}\right|_{(\hat{\mathbf{x}}_k^+, \mathbf{u}_k, \mathbf{0})} = T_s \frac{R}{2} \begin{bmatrix} \cos\hat{\theta}_k^+ & \cos\hat{\theta}_k^+ \\ \sin\hat{\theta}_k^+ & \sin\hat{\theta}_k^+ \\ \dfrac{2}{L} & \dfrac{2}{L} \end{bmatrix} \quad (32)$$

$$H_k = V_k = I \quad (33)$$

### 4.2 Mô phỏng bài toán điều khiển ổn định

Để đánh giá thuật toán đề xuất, chương trình trước hết kiểm tra tính đúng đắn của luật điều khiển (4) trong việc đảm bảo sự ổn định tiệm cận cho hệ robot khi không có trễ. Hình 4 trình bày kết quả mô phỏng điều khiển robot với trạng thái đầu (-3, -3, 0⁰), trạng thái đích (0, 0, 0⁰). Có thể thấy các tọa độ (*x*, *y*) và hướng *θ* của robot tiến về trạng thái đích (0, 0, 0⁰) đồng thời vận tốc dài và vận tốc góc cũng tiến về 0. Hệ thống do đó ổn định tiệm cận. Bây giờ xét hệ thống khi điều khiển phân tán qua mạng. Điều này được mô phỏng bằng việc tạo ra sự trì trễ ở tín hiệu hệ thống. Kết quả mô phỏng luật điều khiển (4) với các tín hiệu điều khiển và phép đo phản hồi bị trì trễ lần lượt 100ms (*n*=10) và 200ms (*m*=20) được thể hiện trong hình 5. Mặc dù tọa độ (*x*, *y*) có xu hướng tiến về (0, 0) nhưng giá trị hướng *θ* lại không hội tụ về 0. Ngoài ra, vận tốc góc ω cũng tiến tới giá trị 1.7rad thay vì 0. Hệ thống do đó không ổn định tiệm cận. Hình 6 trình bày kết quả mô phỏng cho trường hợp trễ ở trên với phần ước lượng trạng thái sử dụng bộ lọc PO-PF. Có thể nhận thấy rằng khi áp dụng bộ lọc PO-PF thì hệ thống lại ổn định tiệm cận trở lại. Hình 7 trình bày kết quả điều khiển ổn định dùng PO-PF cho trường hợp điểm xuất phát (-1.5,-2,π/2), điểm đích (0,0,0$^0$), trễ tín hiệu 150ms (*n*=15) và trễ hệ thống 250ms (*m*=25).

## 5. Thảo luận

Hình 8 so sánh độ lệch giữa tọa độ ước lượng so với tọa độ thực trong hai trường hợp sử dụng bộ lọc Kalman mở rộng và sử dụng bộ lọc PO-PF. Độ lệch ít hơn của bộ lọc PO-PF chứng tỏ hiệu quả của giải thuật cũng như giải thích sự ổn định tiệm cận của hệ thống phân tán qua mạng.

Chúng tôi đã tiến hành mô phỏng với nhiều vị trí xuất phát của robot cũng như với độ trì trễ khác nhau của mạng truyền thông. Trong các trường hợp, bộ lọc PO-PF đều đảm bảo tính ổn định tiệm cận của hệ thống. Trong quá trình mô phỏng, chúng tôi cũng nhận ra rằng thiết lập giá trị ban đầu cho ma trận hiệp phương sai sai số ước lượng *P* đóng vai trò quan trọng cho sự hội tụ của thuật toán. Giá trị *P* thường chọn nhỏ hơn ma trận hiệp phương sai nhiễu phép đo *R* để tránh sự thay đổi đột biến trong các ước lượng đầu tiên khi bắt đầu có dữ liệu đo phản hồi.

## 6. Kết luận

Trong bài báo này, chúng tôi đã đề xuất giải thuật cho phép điều khiển ổn định hệ robot di động phân tán qua mạng máy tính chịu sự tác động của thời gian trễ. Đóng góp chính của bài báo là việc xây dựng một bộ lọc ước lượng trạng thái mới, trên cơ sở lý thuyết về bộ lọc Kalman, cho phép ước lượng và dự đoán trạng thái của hệ thống tại thời điểm hiện tại và tương lai (giới hạn bởi thời gian trễ của tín hiệu điều khiển lối vào) từ dữ liệu phép đo bị trì trễ. Kết quả mô phỏng đã khẳng định tính đúng đắn của thuật toán. Trong nghiên cứu tiếp theo, nhóm tác giả sẽ ứng dụng giải thuật vừa xây dựng vào hệ robot thực đã phát triển tại phòng thí nghiệm. Các vấn đề khác liên quan tới truyền thông qua mạng máy tính như sự mất mát dữ liệu, sai thứ tự gói tin hay băng thông giới hạn cũng sẽ được khảo sát.

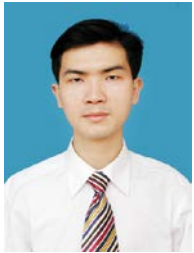

**Phùng Mạnh Dương** nhận bằng cử nhân tại trường Đại học Công nghệ, Đại học Quốc gia Hà Nội năm 2005. Hiện anh là Nghiên cứu sinh tại Khoa Điện tử - Viễn thông, trường Đại học Công Nghệ, Đại Học Quốc Gia Hà Nội. Hướng nghiên cứu chính bao gồm các hệ robot di động phân tán qua mạng máy tính.

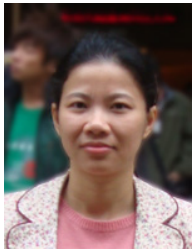

**Nguyễn Thị Thanh Vân** sinh năm 1979. Nhận bằng Cơ điện tử của Viện Công nghệ Châu Á (AIT), Thái Lan năm 2006. Từ năm 2007 đến nay là giảng viên Khoa Điện tử- Viễn thông, Đại học Công nghệ, Đại học Quốc gia Hà Nội. Hướng nghiên cứu chính về các hệ thống điều khiển, điều khiển dẫn đường cho robot di động.

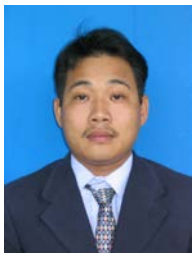

**Trần Thuận Hoàng** sinh năm 1970. Anh nhận bằng thạc sỹ về Đo lường và các hệ thống điều khiển của trường Đại học Bách Khoa Đà Nẵng năm 1998, nhận bằng thạc sỹ Mạng và hệ thống điện năm 2009 của Đại Học Đà Nẵng. Anh hiện là nghiên cứu sinh tại Khoa Điện tử - Viễn thông, trường Đại học Công Nghệ, Đại Học Quốc Gia Hà Nội. Hướng nghiên cứu chính bao gồm tổng hợp các cảm biến cho định vị và dẫn đường robot di động.

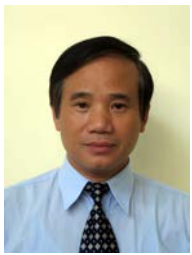

**Trần Quang Vinh** bảo vệ tiến sĩ Vật lý Vô tuyến điện tại ĐH Quốc gia Hà nội trên cơ sở các nghiên cứu thực nghiệm tại Đại học Tổng hợp Kỹ thuật TU Wien (Áo) năm 2001. Hiện là Phó giáo sư, Chủ nhiệm Bộ môn Điện tử và Kỹ thuật máy tính, Trưởng phòng thí nghiệm Các hệ tích hợp thông minh (SIS) tại trường ĐH Công nghệ. Hướng chuyên môn quan tâm hiện nay: Đo lường và điều khiển dùng vi tính và vi xử lý cho các lĩnh vực: vật lý, hóa học, môi trường, y-sinh, nhà thông minh; Điều khiển tự động và robot thông minh (robot di động tự trị, robot nối mạng); Thiết kế chip điện tử tích hợp cỡ lớn VLSI, FPGA, ASIC.